\pdfoutput=1
\documentclass[a4paper]{article}

\usepackage[OT1]{fontenc}
\usepackage{spconf}
\usepackage{hyperref}
\hypersetup{
  colorlinks,
  linkcolor={black},
  citecolor={black},
  urlcolor={black}
}
\usepackage{booktabs}
\usepackage{tikz}
\usepackage{pgfplots}
\usepackage{microtype}
\usepackage{mathtools}

\usepackage{siunitx}
\providecommand{\qty}[2]{\SI{#1}{#2}}
\providecommand{\unit}[1]{\si{#1}}

\newcommand{\Ad}{\textsf{Ad}}
\newcommand{\Le}{\textsf{Le}}
\newcommand{\AdLo}{\textsf{AdA}}
\newcommand{\LeLo}{\textsf{LeA}}
\newcommand{\groundt}{\textsf{GT}}

\newlength{\floatcorrection}
\newcommand\trimfloat{\vspace{-\floatcorrection}}

\definecolor{mpl0}{RGB}{31,119,180}
\definecolor{mpl1}{RGB}{255,127,14}
\definecolor{mpl2}{RGB}{44,160,44}
\definecolor{mpl3}{RGB}{214,39,40}
\definecolor{mpl4}{RGB}{148,103,189}
\definecolor{mpl5}{RGB}{176,176,176}

\title{Analysis and transformation of voice level in singing voice}
\name{%
  \href{mailto:frederik.bous@ircam.fr}{Frederik Bous},
  Axel Roebel%
  \thanks{%
    This work has been funded partly by the ANR project ARS (ANR-19-CE38-0001-01).
    This work was performed using HPC resources from GENCI-IDRIS
    (Grants 2020-AD011011378 and 2021-AD011011177).
  }
}
\address{%
  UMR 9912 STMS, %\\%
  IRCAM, Sorbonne Universit{\'e}, CNRS, % \\%
  Paris, France%
}

\begin{document}
  \ninept
  \maketitle
  \begin{abstract}
    We introduce a neural auto-encoder that transforms
the musical dynamic in recordings of singing voice
via changes in voice level.
Since most recordings of singing voice are not annotated
with voice level
we propose a means to estimate the voice level
from the signal's timbre
using a neural voice level estimator.
We introduce the \emph{recording factor}
that relates the voice level to the recorded signal power
as a proportionality constant.
This unknown constant depends on the recording conditions
and the post-processing
and may thus be different for each recording
(but is constant across each recording).
We provide two approaches to estimate the voice level
without knowing the recording factor.
The unknown recording factor can either be learned
alongside the weights of the voice level estimator,
or a special loss function
based on the scalar product
can be used to only match the contour
of the recorded signal's power.
The voice level models are used
to condition a previously introduced bottleneck auto-encoder
that disentangles its input, the mel-spectrogram,
from the voice level.
We evaluate the voice level models
on recordings annotated with musical dynamic
and by their ability to provide useful information
to the auto-encoder.
A perceptive test is carried out
that evaluates the perceived change in voice level
in transformed recordings
and the synthesis quality.
The perceptive test confirms
that changing the conditional input
changes the perceived voice level accordingly
thus suggesting
that the proposed voice level models
encode information about the true voice level.

  \end{abstract}

  \begin{keywords}
    Voice transformation, voice conversion, auto-encoder, voice level%
    \vspace*{-2ex}
  \end{keywords}

  \section{Introduction}
  Voice level is the power with which human voice is produced.
  For singing, the voice level is strongly related to the musical dynamic
  and therefore an important medium to carry musical expression
  \cite{umbert2015expression}.
  In music the dynamic refers to how loud an instrument is played.
  Singing voice as a musical instrument
  inherits this quality
  and professional singers can sing a continuous spectrum of dynamics
  by adjusting the voice level.
  Similarly in speech most speakers can adapt the voice level
  to various situations between speaking to a crowd of hundreds of people
  in an open field
  and exchanging information with their neighbour in a quiet library
  \cite{traunmuller2000acoustic}.

  For both, singing and speech,
  not only the signal's power changes with the change in voice level
  but a wide range of voice properties as well,
  such that most people can easily distinguish
  soft speech close to the ear
  from someone shouting from far away,
  even though the signal's power might be the same
  at the listeners ear \cite{chowning2000digital}.
  In fact the perceived loudness of sounds
  does only marginally depend
  on the signal power at the listener
  as our brain compensates for attenuations
  due to distance to the source \cite{chowning2000digital}.

  Voice level can be derived from the intensity measured
  using a calibrated \emph{sound pressure level} (SPL) meter
  at a fixed distance to the singer
  \cite{traunmuller2000acoustic}
  and the relationship between voice level
  and other voice parameters have been investigated
  in various studies.
  In \cite{holmberg1988glottal}
  glottal inverse filtering \cite{alku2011glottal}
  is used to analyse the difference between soft, normal and loud voice
  in the glottal waveform.
  The relationships between open quotient, $f_0$, lung pressure
  and sound pressure level
  are investigated in \cite{titze1992vocal}.
  In \cite{traunmuller2000acoustic}
  the relationship between vocal effort
  and the articulation, $f_0$, creaky voice and formant positions
  is studied
  and the distance between speaker and addressee is used
  to vary the vocal effort / voice intensity of the participants.
  Special attention to the open quotient
  is given by the study in \cite{henrich2005glottal}.

  These studies have paved the way
  for glottal pulse models \cite{doval2006spectrum}
  which can be used to modify pulse parameters
  as a function of the voice level
  \cite{roebel2012analysis,huber2015glottal}.
  However, the practical application of these models is limited
  as robustly obtaining glottal pulse parameters from voice recordings
  remains challenging.
  In \cite{molina2014parametric}
  voice level changes were done by
  shifting the formant positions in vowels
  according to the statistics computed on a singing database.
  Related to the voice level in voice is the roughness
  as rough voice generally tents to
  occur in voice with very high voice level.
  Rough voice is, however, a different singing style
  that can happen in quiet singing as well.
  Some work has been done
  to simulate the effect or rough voice
  in singing synthesis
  \cite{gentilucci2018vocal,gentilucci2019composing}.

  In this paper we aim to use neural networks
  to transform the voice level of singing voice recordings.
  Based on the observations above
  the voice level manifests itself heavily in the timbre.
  Thus it does not suffice to simply rescale the original signal.
  Numerous datasets with singing voice exist,
  however only a small fraction contains voice level annotations.
  Therefore, in this paper we provide a method
  to train a neural network to extract voice level
  without the need for explicitly annotated voice level.
  The estimated voice level can be calibrated
  using any of the recording conditions
  present in the database of singing recordings
  as a reference condition.

  Thus, the contributions of this work are the following.
  (1.) We present an unsupervised algorithm
  that allows training a deep neural network
  to predict the voice level
  without the need to create annotated data.
  (2.) We use this voice level measure to train
  a bottleneck auto-encoder \cite{qian2019autovc,qian2020f0}
  to transform the voice level in recordings of singing voice.
  The remaining paper is structured as follows:
  In Section~\ref{sec:analysis}
  we introduce two ways
  to estimate the voice level
  from voice recordings
  where no voice level annotation exists.
  In Section~\ref{sec:transformation}
  we present our adaptions of \cite{bous2022bottleneck}
  to include the voice level
  as a controllable parameter.
  We will explain our experimental validation
  in Section~\ref{sec:experiments}
  and present and discuss the results in Section~\ref{sec:results}.
  Finally we will see a short summary and an outlook
  in Section~\ref{sec:conclusions}.

  \vspace*{-2ex}
  \section{Extracting voice level from audio recordings}
  \label{sec:analysis}
  \vspace*{-1ex}
  \newcommand{\level}{\ensuremath{l}}
  \newcommand{\power}{\ensuremath{p}}
  \newcommand{\refpt}{\ensuremath{\mathbf{x}}}

  Audio recordings
  are not calibrated measurements.
  The goal of a recording
  is to produce a signal
  that when played on a speaker
  will create sound waves
  that sound like the original source.
  The scaling of these signals is irrelevant
  as it is expected to be adjusted
  by the consumer or the sound engineer that further processes the sound.
  Thus, each recording
  has a different relationship between the source's power
  and the recorded signal's power as
  microphones have different directivities and transfer functions
  and the signal gain is adjusted
  to minimise quantisation noise
  when digitalising the microphone signal.

  Therefore, without explicit annotation
  it is impossible to infer the voice level
  from an audio recording's power.
  Still, if we look at the speech production mechanism,
  we notice that humans cannot increase the voice level
  without changing other properties of their voice
  \cite{traunmuller2000acoustic,holmberg1988glottal,titze1992vocal,henrich2005glottal}.
  Thus we can infer the voice level from the signal's timbre.

  \subsection{Learned recording factor}
  \label{ssec:learnedfactor}

  \newcommand{\speakerid}{\ensuremath{\mathsf{s}}}
  \newcommand{\speakerids}{\ensuremath{\mathsf{S}}}
  \newcommand{\weights}{\ensuremath{\theta}}
  \newcommand{\argmin}{\ensuremath{\operatorname*{argmin}}}
  \newcommand{\distance}{\operatorname{dist}}
  \newcommand{\mel}{\ensuremath{M}}
  \newcommand{\expectation}{\mathbf{E}}
  \newcommand{\normsq}[1]{\ensuremath{\left\lVert#1\right\rVert_2^2}}
  \newcommand{\networkfn}[2][\weights]{N_{#1}\left({#2}\right)}
  \newcommand{\normalise}{\ensuremath{\operatorname{n}}}

  Assuming all audio files of a dataset
  have been recorded under the same conditions,
  (same microphone, pre-gain, spatial positioning of speaker and microphone, etc.)
  with the same post-processing applied to them
  (in particular with same normalisation factor)
  we can assume that
  the power contour of the recorded signal $\power$
  is proportional to the voice level $\level$:
  \begin{align}
    \label{eq:spkfact}
    \power = a_r \, \level
  \end{align}
  with \power\ and \level\ being time varying sequences here
  and with the proportionality factor $a_r$
  which we shall call \emph{recording factor}
  as it captures the effect of the recording conditions.
  For multi speaker databases
  it is highly likely that
  some of these assumptions are violated;
  however, it is not unlikely
  that these assumptions still hold
  for all files of a fixed speaker $\speakerid$.
  Therefore we get the relationship
  \begin{align}
    \power = a_r^{\speakerid} \, \level
  \end{align}
  for all files generated by a speaker $\speakerid$
  with a different $a_r^{\speakerid}$ for each speaker.

  The speaker dependent recording factor $a_r^{\speakerid}$
  can be learned alongside the weights $\weights$
  of a neural network $N_{\weights}$.
  As we aim to learn the voice level contour \level\ 
  from the spectral properties of the signal
  we have to prevent the network from using the signal power.
  This can be done by frame-wise normalisation
  thus removing the power contour from the input.

  Using an $L_2$ error,
  we get the following error function
  \begin{align}
    \normsq{\power - a_r^{\speakerid} \, \networkfn{\normalise(x)}}
  \end{align}
  with frame-wise normalisation \normalise.

  The resulting neural network $N_{\weights}$
  allows estimating the voice level
  from the normalised signal:
  During inference we use the same recording factor for all recordings,
  in which case we can compare the recordings
  as if they were made under the same recording conditions.
  To calibrate the estimations to a specific recording
  we use the recording factor of that recording.

  \subsection{Adaptive recording factor}
  \label{ssec:adaptive}

  \newcommand{\target}{\ensuremath{\power}}
  \newcommand{\mout}{\ensuremath{q}}
  \newcommand{\scalarprod}[2]{\ensuremath{#1\cdot#2}}
  \newcommand{\scploss}{\ensuremath{e_{\text{scp}}}}
  \newcommand{\invarloss}{\ensuremath{e_{\text{invar}}}}

  The assumptions from Section~\ref{ssec:learnedfactor}
  require the files from the training dataset
  to be grouped by same recording conditions.
  This works well if the number of speakers is small
  and we can be sure that the files have not been normalised separately.
  However, for many databases
  we don't know what kind of post-processing has been performed
  or whether the samples for a fixed speaker
  have been created over multiple recording sessions
  with slightly different conditions.
  In this case we would have to assign a different recording factor to each file.
  With the previous approach
  this causes problems as a gradient exists for a specific recording factor
  only if a sample associated with this recording factor is present in the training batch.
  Thus recording factors with a small percentage of associated files in the dataset
  are learned very slowly as they are updated rarely.

  For the case where we cannot group the files
  into a reasonable amount of classes
  we propose an adaptive recording factor:
  Let $\mout := \networkfn{\normalise(x)}$ be the output of the neural network
  and \target\ the power curve associated with the input sample.
  Again, we assume \eqref{eq:spkfact},
  this time with a different recording factor $a$
  for each training sample.
  Prior to calculating the loss
  we choose $a$ such that the $L_2$ error
  \begin{align}
    \label{eq:l2err}
    e_a = \normsq{\target - a \mout} = \sum_t (\target_t - a \mout_t)^2
  \end{align}
  is minimal for each training sample:
  \begin{align}
    \hat{a} = \argmin_{a} e_a
  \end{align}

  For a fixed pair of target and estimate $(\target, \mout)$
  there exists an analytic optimal solution for~$a$:
  \begin{align}
    \label{eq:ahat}
    \hat{a}
      = \frac{\scalarprod\target\mout}{\normsq{\mout}}
      = \frac{\sum_t \target_t \mout_t}{\sum_{\tau} \mout_{\tau}^2}
  \end{align}
  where $\cdot$ denotes the scalar product.
  Combining \eqref{eq:l2err} and \eqref{eq:ahat}
  and normalising $e_{\hat{a}}$ by $\normsq{\target}$
  yields the scalar product loss:
  \begin{align}
    \scploss := \frac{e_{\hat{a}}}{\normsq{\target}}
      = 1 - \frac{(\scalarprod\target\mout)^2}{\normsq{\target}\normsq{\mout}}
      = 1 - \left(\scalarprod{\bar{\target}}{\overline{\mout}}\right)^2
      % &= 1 - \cos^2\angle(\target,\mout) \\
      % &= \sin^2\angle(\target,\mout)             \text{,}
  \end{align}
  with $\bar{x} := x / \lVert x \rVert$ denoting the unit vector in direction of $x$.

  With the same reasoning as before
  the input signal has to be normalised
  to remove information of the signal energy.

  To calibrate the estimation during inference
  to the recording environment of a specific recording,
  we can calculate the recording factor according to~\eqref{eq:ahat}
  and rescale the network output by that factor.

  \section{Proposed voice level transformations}
  \label{sec:transformation}

  Having a method to infer the voice level from audio recordings
  could be useful for many applications
  and in different disciplines.
  In this publication we focus our attention
  on transforming the voice level in singing voice.
  We can use a bottleneck auto-encoder \cite{qian2019autovc}
  to disentangle the voice level
  from the mel-spectrogram of singing voice recordings.
  We extend the architecture of \cite{bous2022bottleneck}
  to additionally include the voice level
  as conditional input.

  We can use this auto-encoder to validate the proposed voice level estimator
  that was introduced in Section~\ref{sec:analysis}
  and show that it really represents
  the (perceived) voice level.
  If the transformed recordings
  are perceived to have been sung with a voice level
  that is coherent with the intended transformation
  (louder, less loud),
  we can conclude that the proposed voice level estimators
  in fact encode the voice level.

  \section{Experiments}
  \label{sec:experiments}

  \subsection{Data}
  All experiments are trained on the same dataset as
  \cite{bous2022bottleneck},
  which is a combined dataset of 
  % CREL
  CREL Research Database (SVDB) \cite{tsirulnik2019singing},
  % NUS_SMC
  NUS sung and spoken lyrics corpus \cite{duan2013nus},
  % Dimitrios
  from the i-Treasures Intangible Cultural Heritage dataset
  \cite{grammalidis2016treasures}
  % PJS
  PJS phoneme-balanced Japanese singing-voice corpus \cite{koguchi2020pjs},
  % JVS
  JVS-MuSiC \cite{tamaru2020jvs},
  % Tohoku
  Tohoku Kiritan and Itako singing database \cite{ogawa2021tohoku},
  % VocalSet
  VocalSet: A singing voice dataset \cite{wilkins2018vocalset},
  % DeepFarinelli, ISiS
  as well as singing recordings from our internal singing databases
  used for the IRCAM Singing Synthesizer \cite{ardaillon2017synthesis}
  and other projects.

  \subsection{Architecture}
  We train two voice level estimators,
  one with the learned recording factor
  from Section~\ref{ssec:learnedfactor} (\Le)
  and the other with the adaptive recording factor strategy
  from Section~\ref{ssec:adaptive} (\Ad).
  As input to the estimators we choose the signal's mel-spectrograms
  to provide a concise representation of the spectral properties
  to the voice level estimators
  and to match the input data to the auto-encoders
  that will be conditioned on these voice level estimators.
  Thus we use the same analysis parameters as in \cite{bous2022bottleneck}
  to generate the mel-spectrograms.
  We experimented with different ways to estimate the signal power
  including the short-term-energy
  and perceptive loudness measures
  and found that we achieve the best results
  using the loudness measure from \cite{glasberg2002model}.

  Both networks, \Le and \Ad, have the same architecture.
  The networks are simple convolutional feed-forward networks with 10 layers.
  Convolutions are 1d, treating the frequency bins as features.
  The number of filters is 100 in most of the layers,
  except the first, which has 80,
  the second to last, which has 50,
  and the last which has 1.
  The filter size is 3 in the first two layers
  and 1 elsewhere.
  With a step size of \qty{12.5}{\milli\second} per mel-frame
  the voice level estimator has thus a receptive field
  of 5 frames or \qty{50}{\milli\second}.
  The mel-spectrograms are normalised frame-wise
  when fed as input to the voice level estimators.

  We train the models with a batch size of 256 training samples
  of 80 mel-frames (or \qty{1}{\second}) each.
  The models are trained for 500\,k updates
  using the adam \cite{kingma2014adam} optimiser
  ($\beta_1 = 0.9$, $\beta_2 = 0.999$)
  with an initial learning rate of \num{1e-4}.
  The learning rate is reduced by a factor of $\sqrt[4]{0.1}$
  if the validation loss does not decrease
  for a period of 16\,k updates,
  with a minimum learning rate of \num{1e-6}.

  \subsection{Auto-encoders}
  We train two auto-encoder configurations,
  one for each voice level estimator.
  For the auto-encoder we use almost the same architecture
  as in \cite{bous2022bottleneck}
  only with the conditional input changed,
  as we add the voice level to the list of conditional inputs.
  Thus the auto-encoders are conditioned on the voice level,
  the $f_0$ and the voiced-unvoiced mask.
  We use the $f_0$ model from \cite{ardaillon2019fully}
  and the mel-inverter from \cite{roebel2022neural}.

  \subsection{Evaluation methods}
  Since the dataset that we used to train our models
  does not include annotations for the voice level,
  we cannot evaluate the models directly with a ground truth.
  The hypothesis of this paper is
  that we can extract meaningful information
  from the spectral properties of the mel-spectrogram
  about the perceived voice level.
  Thus, it suffices to show
  that the information extracted by our voice level estimators
  reflects the perceived voice level.

  We use recordings that are labelled into different dynamics
  to investigate the relationship between dynamic and the proposed voice level estimator.
  Labelled data was obtained in \cite{ardaillon2017synthesis}
  by asking a singer to sing one note
  from all combinations of the dynamics
  pp, mp, mf, f and ff and all French vowels
  on the same pitch.
  Thus we can create histograms over the estimated voice level
  for different dynamics and vowels.
  If the proposed method reflects the true voice level,
  lower dynamics should correspond to lower values in voice level.

  Furthermore, we claim that the proposed voice level estimate
  is useful for changing how loud a phrase has been sung.
  Thus, if the proposed auto-encoder indeed succeeds in changing the signals properties
  in a way that it is perceived as sung with the desired voice level,
  we can conclude that the proposed voice level estimators
  indeed encode the voice level.
  To this end we asked
  40 participants
  in a perceptive online study
  to rate the perceived voice level change
  of the transformed audio.
  Participants were presented pairs of audio
  where for each pair
  both files were generated using one of the auto-encoders
  and where in one file the voice level was changed
  while for the other the voice level was kept the same.
  Participants were the asked to rate
  which recording sounded
  as if it was sung with a stronger / louder voice
  and could give an answer of -2, -1, 0, 1 or 2.
  The order within each pair
  and the overall order of the pairs were randomised
  and the volume of each of the files
  was normalised to the same average loudness
  according to the loudness model of \cite{glasberg2002model}.
  In a second test, we asked 26 participants
  to rate the audio samples for their audio quality
  and computed a mean-opinion-score for each of the voice level changes
  of both models and the ground truth reference.

  The auto-encoder's precision is evaluated
  by measuring the average difference between requested voice level
  and voice level measured in the synthesised mel-spectrograms
  for various voice level changes.
  The results are given in Table~\ref{tab:accuracy}.
  The samples used for the perceptive test
  are available on our website%
  \footnote{%
    \href{%
      http://recherche.ircam.fr/anasyn/bous/aeint2022%
    }{%
      recherche.ircam.fr/anasyn/bous/aeint2022%
    }
  }.

  \section{Results}
  \label{sec:results}

  \subsection{Classification of dynamic}

  \def\figH{3cm}
  \newcommand\createFirstAxis[1]{
    \begin{tikzpicture}
      \begin{axis}[
        height=\figH,
        width=\linewidth,
        ylabel={/#1/},
        yticklabel style={
          /pgf/number format/fixed,
        },
        xticklabels={,,},
        xmajorgrids=true,
        ymajorgrids=true,
        xmin=-33,
        xmax=-3,
        legend columns=5,
        legend style={
          anchor=south,
          at={(0.5,1.10)},
        }
      ]
        \addplot [densely dotted, mpl0] table[y=pp] {fig/dyn_dist_#1.dat};
        \addplot [dashed,mpl1] table[y=mp] {fig/dyn_dist_#1.dat};
        \addplot [dashdotted,mpl2] table[y=mf] {fig/dyn_dist_#1.dat};
        \addplot [densely dashdotdotted,mpl3] table[y=f] {fig/dyn_dist_#1.dat};
        \addplot [thin,mpl4] table[y=ff] {fig/dyn_dist_#1.dat};
        \legend{pp,mp,mf,f,ff}
      \end{axis}
    \end{tikzpicture}
  }
  \newcommand\createAxis[1]{
    \begin{tikzpicture}
      \begin{axis}[
        height=\figH,
        width=\linewidth,
        ylabel={/#1/},
        yticklabel style={
          /pgf/number format/fixed,
        },
        xticklabels={,,},
        xmajorgrids=true,
        ymajorgrids=true,
        xmin=-33,
        xmax=-3,
      ]
        \addplot [densely dotted, mpl0] table[y=pp] {fig/dyn_dist_#1.dat};
        \addplot [dashed,mpl1] table[y=mp] {fig/dyn_dist_#1.dat};
        \addplot [dashdotted,mpl2] table[y=mf] {fig/dyn_dist_#1.dat};
        \addplot [densely dashdotdotted,mpl3] table[y=f] {fig/dyn_dist_#1.dat};
        \addplot [thin,mpl4] table[y=ff] {fig/dyn_dist_#1.dat};
      \end{axis}
    \end{tikzpicture}
  }
  \newcommand\createFullAxis[1]{
    \begin{tikzpicture}
      \begin{axis}[
        height=\figH,
        width=\linewidth,
        xlabel={Voice level [dB]},
        ylabel={#1},
        yticklabel style={
          /pgf/number format/fixed,
        },
        xmajorgrids=true,
        ymajorgrids=true,
        xmin=-33,
        xmax=-3,
      ]
        \addplot [densely dotted, mpl0] table[y=pp] {fig/dyn_dist_#1.dat};
        \addplot [dashed,mpl1] table[y=mp] {fig/dyn_dist_#1.dat};
        \addplot [dashdotted,mpl2] table[y=mf] {fig/dyn_dist_#1.dat};
        \addplot [densely dashdotdotted,mpl3] table[y=f] {fig/dyn_dist_#1.dat};
        \addplot [thin,mpl4] table[y=ff] {fig/dyn_dist_#1.dat};
      \end{axis}
    \end{tikzpicture}
  }

  \newcommand\figDyndist{
    \begin{figure}[tb]
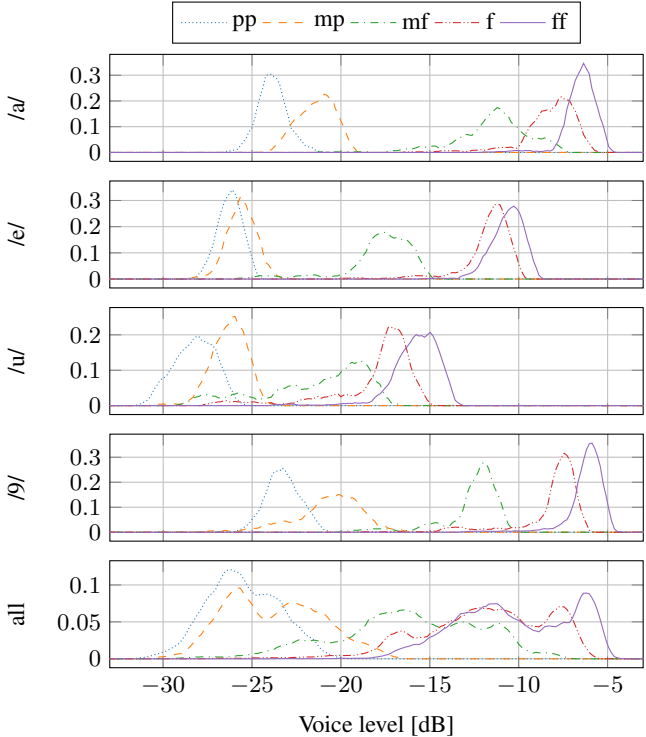

      \createFirstAxis{a}
      \createAxis{e}
      \createAxis{u}
      \createAxis{9}
      \createFullAxis{all}
      \caption{
        Estimated voice level values
        for recordings in different dynamics
        plotted for the voice level model \Ad.
        The graphs show smoothed relative histograms
        using a Gaussian kernel with $\sigma=\qty{0.31}{\decibel}$.
        Only frames inside stable phonation are used,
        unvoiced frames and frames close (\qty{50}{\milli\second})
        to a voiced / unvoiced boundary are ignored.
        We show histograms for four selected vowels
        (out of a total of 15 vowels in the French language)
        and the average over all vowels (bottom).
        All graphs share the same horizontal axis.
        Phoneme annotations are in X-SAMPA.
      }
      \label{fig:dyndist}
      \trimfloat
    \end{figure}
  }

  Figure~\ref{fig:dyndist} visualises the relationship
  between estimated voice level of \Ad\ 
  and the symbolic dynamic annotation
  of the recordings used for this investigation.
  Similar histograms are produced by model \Le.
  First, we observe that louder dynamics
  produce higher voice level values
  so we can conclude that the voice level
  does indeed include information
  about how loud a note has been sung.
  If we look at the individual phonemes,
  we see that the dynamics create different clusters
  with increasing mean
  though for some vowels some dynamics have rather large overlap,
  e.\,g.\ for /e/ the pp and mp are largely overlapping:
  This is because
  we see the histogram of the voice level for each frame
  which naturally varies over time for the same dynamic
  for instance as a side effect of vibrato.
  Thus while some selected frames from different dynamics may be equally loud,
  the overall notes still have different loudness
  and can be measured as averages over larger durations.
  Furthermore, dynamics are subjective and a matter of interpretation.
  It could be that this particular singer
  does not make such a big difference between pp and mp for an /e/
  in terms of voice level.
  \figDyndist

  Comparing the histograms from different vowels
  we observe that the voice level
  can be very different for different phonemes
  under the same dynamic.
  This is because not all vowels are in fact equally loud.
  The vowel /a/,
  being an open vowel,
  can be sung much louder than /u/,
  being a close vowel.
  Thus for the same dynamic
  we can expect the resulting voice level to be different
  and it is no surprise
  that the estimated voice level for /a/
  is much higher than for /u/.
  This explains why,
  if we average over all vowels,
  adjacent dynamics overlap ---
  because the variations in voice level due to the phoneme
  are on the same scale as the variations due to the dynamic.
  Nevertheless, we can still very clearly distinguish
  between loud dynamics (f, ff) and quiet dynamics (pp, mp).

  \subsection{Accuracies}

  \begin{table}[tb]
    \footnotesize
    \begin{tabular}{lrrrrr}
      \toprule
      & \qty{-10}{\decibel}
      & \qty{-6}{\decibel}
      & \qty{0}{\decibel}
      & \qty{+6}{\decibel}
      & \qty{+10}{\decibel} \\
      \midrule
      \AdLo & 2.42 & 1.50 & 0.86 & 1.94 & 2.50\\
\LeLo & 2.71 & 1.75 & 0.81 & 2.99 & 4.68\\
      \bottomrule
    \end{tabular}
    \caption{
      Transformation precision \unit{[\dB]}
      of the auto-encoders.
      Errors are calculated between target voice level
      (as given to the auto-encoder)
      and voice level of the output
      (the voice level estimator applied to the transformed mel-spectrogram)
      for various voice level changes.
      The error is calculated on a logarithmic scale
      to provide meaningful values.
    }
    \label{tab:accuracy}
    \trimfloat
  \end{table}

  From Table~\ref{tab:accuracy} we can see
  that the auto-encoders can use the information
  provided by the voice level estimator
  and successfully disentangle the voice level
  from the mel-spectrograms.
  The auto-encoder with adapted recording factor (\AdLo)
  performs with higher precision
  than the auto-encoder with learned recording factor (\LeLo).
  This indicates that the adapted recording factor model (\Ad)
  is more robust than the learned recording factor model (\Le)
  as some of the assumptions we made in Section~\ref{ssec:learnedfactor}
  might in fact be violated in the training dataset
  which caused the voice level estimator \Le\ to provide inconsistent information.

  \subsection{Perceptive test}

  \begin{table}[tb]
    \footnotesize
    \resizebox{\linewidth}{!}{
      \addtolength{\tabcolsep}{-2.5pt}
      \begin{tabular}{lcccc}
        \toprule
        \textbf{Voice level} & \qty{-10}{\decibel} & \qty{-6}{\decibel} & \qty{6}{\decibel} & \qty{10}{\decibel} \\
        \midrule
        \AdLo & $-1.57 \pm 0.32$ & $-0.84 \pm 0.20$ & $0.55 \pm 0.20$ & $0.08 \pm 0.32$\\
\LeLo & $-1.29 \pm 0.33$ & $-0.74 \pm 0.22$ & $0.69 \pm 0.18$ & $0.83 \pm 0.22$\\
        \bottomrule
      \end{tabular}
    }
    \resizebox{\linewidth}{!}{
      \addtolength{\tabcolsep}{-2.5pt}
      \footnotesize
      \begin{tabular}{lccccc}
        \toprule
        \textbf{Quality} & \qty{-10}{\decibel} & \qty{-6}{\decibel} & \qty{0}{\decibel} & \qty{6}{\decibel} & \qty{10}{\decibel} \\
        \midrule
        \groundt & & & $4.64 \pm 0.12$ & & \\
\AdLo & $2.12 \pm 0.49$ & $3.32 \pm 0.37$ & $3.82 \pm 0.25$ & $3.55 \pm 0.31$ & $2.78 \pm 0.37$\\
\LeLo & $2.61 \pm 0.44$ & $3.40 \pm 0.39$ & $3.80 \pm 0.26$ & $3.55 \pm 0.30$ & $3.55 \pm 0.39$\\
        \bottomrule
      \end{tabular}
    }
    \caption{
      Results of the perceptive test
      for relative voice level
      (subjective scale from $-2$ to $2$)
      and audio quality
      (subjective scale from $1$ to $5$)
      with 95\% confidence intervals.
    }
    \label{tab:perceptive_test}
    \trimfloat
  \end{table}

  Table~\ref{tab:perceptive_test}
  summarises the results from the perceptive test.
  From the upper table
  we can see, that both models
  are able to create a noticeable change in voice level
  for small changes in the target voice level.
  For the auto-encoder with adaptive recording factor (\AdLo)
  we observe that it has trouble
  creating convincing results for high increases in voice level.
  Investigating the files that were rated in the opposite direction
  we noticed that for those files the auto-encoder
  introduced significant amounts of artefacts
  which seemed to have created the opposite
  of the desired effect.
  The perceptive test suggests
  that the auto-encoders have a more noticeable impact
  when decreasing the voice level,
  which can be seen for both models
  and for all amounts of change.
  The auto-encoder with adaptive recording factor, \AdLo,
  seems to create a stronger effect
  when decreasing the voice level
  than the auto-encoder with learned recording factor, \LeLo.
  For increasing the voice level
  \LeLo\ seems to be better suited than \AdLo.

  The quality ratings are given in the lower table of Table~\ref{tab:perceptive_test}
  For self-reconstruction and small amounts of voice level change
  both models, \AdLo\ and \LeLo,
  seem to work equally well.
  For large changes in voice level
  the auto-encoder with learned recording factor (\LeLo)
  outperforms the auto-encoder with adaptive recording factor (\AdLo)
  by a margin.
  For increases in voice level
  \LeLo\ is able to hold its level of quality
  even for an increase of \qty{10}{\dB}.
  On the other hand \LeLo\ does make a larger error
  in Table~\ref{tab:accuracy}.
  For decrease in voice level
  both auto-encoder models
  suffer strong degradation in quality
  although both models were successfully able
  to convince the participants
  that the recordings had much less voice level.
  Listening to these samples reveals
  that the auto-encoders increase the background noise significantly.
  Since our mel-inverter does not handle synthetic noise well,
  the overall audio quality is poor in these cases
  although the conversion itself is realistic.

  From the test results we can conclude
  that the given auto-encoders
  were able to change
  the perceived voice level in singing voice,
  and therefore the voice level estimators
  capture the information about the true voice level.

  \section{Conclusions}
  \label{sec:conclusions}

  We have introduced a method
  to estimate the voice level
  from recordings with unknown amplification factors
  (recording factor).
  Two variants to overcome the missing the recording factor
  have been proposed:
  either by learning the unknown recording factor
  alongside the weights of the neural network (\Le)
  or by adjusting the loss function
  to remove scaling and only compare the contours
  of the network's output and the associated signal power (\Ad).
  These voice level estimators have been used
  to condition a bottleneck auto-encoder
  to disentangle the voice level from mel-spectrograms.
  We have shown that both models produce consistent values
  and can produce the effect of changed voice level
  on singing recordings in most cases
  and with acceptable quality.

  While the proposed auto-encoders produce
  a noticeable change in voice level,
  the audio quality is still significantly lower than real recordings
  especially when increasing the voice level.
  Consequently this first publication
  on neural voice level transformation
  has to be seen as a proof-of-concept
  rather than a well-polished system.
  Improvements to the audio quality are required
  for this method to be used in actual musical production.
  Nevertheless, the estimation method for the voice level
  opens new ways of voice classification, analysis and transformation.

  \bibliographystyle{IEEEtran}
  \bibliography{refs}

\begin{thebibliography}{10}
\providecommand{\url}[1]{#1}
\csname url@samestyle\endcsname
\providecommand{\newblock}{\relax}
\providecommand{\bibinfo}[2]{#2}
\providecommand{\BIBentrySTDinterwordspacing}{\spaceskip=0pt\relax}
\providecommand{\BIBentryALTinterwordstretchfactor}{4}
\providecommand{\BIBentryALTinterwordspacing}{\spaceskip=\fontdimen2\font plus
\BIBentryALTinterwordstretchfactor\fontdimen3\font minus
  \fontdimen4\font\relax}
\providecommand{\BIBforeignlanguage}[2]{{%
\expandafter\ifx\csname l@#1\endcsname\relax
\typeout{** WARNING: IEEEtran.bst: No hyphenation pattern has been}%
\typeout{** loaded for the language `#1'. Using the pattern for}%
\typeout{** the default language instead.}%
\else
\language=\csname l@#1\endcsname
\fi
#2}}
\providecommand{\BIBdecl}{\relax}
\BIBdecl

\bibitem{umbert2015expression}
M.~Umbert, J.~Bonada, M.~Goto, T.~Nakano, and J.~Sundberg, ``Expression control
  in singing voice synthesis: Features, approaches, evaluation, and
  challenges,'' \emph{IEEE Signal Processing Magazine}, vol.~32, no.~6, pp.
  55--73, 2015.

\bibitem{traunmuller2000acoustic}
H.~Traunm{\"u}ller and A.~Eriksson, ``Acoustic effects of variation in vocal
  effort by men, women, and children,'' \emph{The Journal of the Acoustical
  Society of America}, vol. 107, no.~6, pp. 3438--3451, 2000.

\bibitem{chowning2000digital}
J.~M. Chowning, ``Digital sound synthesis, acoustics and perception: A rich
  intersection,'' in \emph{COST G-6 Conference on Digital Audio Effects
  (DAFX-00)}, 2000.

\bibitem{holmberg1988glottal}
E.~B. Holmberg, R.~E. Hillman, and J.~S. Perkell, ``Glottal airflow and
  transglottal air pressure measurements for male and female speakers in soft,
  normal, and loud voice,'' \emph{The Journal of the Acoustical Society of
  America}, vol.~84, no.~2, pp. 511--529, 1988.

\bibitem{alku2011glottal}
P.~Alku, ``Glottal inverse filtering analysis of human voice production—a
  review of estimation and parameterization methods of the glottal excitation
  and their applications,'' \emph{Sadhana}, vol.~36, no.~5, pp. 623--650, 2011.

\bibitem{titze1992vocal}
I.~R. Titze and J.~Sundberg, ``Vocal intensity in speakers and singers,''
  \emph{the Journal of the Acoustical Society of America}, vol.~91, no.~5, pp.
  2936--2946, 1992.

\bibitem{henrich2005glottal}
N.~Henrich, C.~d’Alessandro, B.~Doval, and M.~Castellengo, ``Glottal open
  quotient in singing: Measurements and correlation with laryngeal mechanisms,
  vocal intensity, and fundamental frequency,'' \emph{The Journal of the
  Acoustical Society of America}, vol. 117, no.~3, pp. 1417--1430, 2005.

\bibitem{doval2006spectrum}
B.~Doval, C.~d'Alessandro, and N.~Henrich, ``The spectrum of glottal flow
  models,'' \emph{Acta acustica united with acustica}, vol.~92, no.~6, pp.
  1026--1046, 2006.

\bibitem{roebel2012analysis}
A.~Roebel, S.~Huber, X.~Rodet, and G.~Degottex, ``Analysis and modification of
  excitation source characteristics for singing voice synthesis,'' in
  \emph{2012 IEEE International Conference on Acoustics, Speech and Signal
  Processing (ICASSP)}.\hskip 1em plus 0.5em minus 0.4em\relax IEEE, 2012, pp.
  5381--5384.

\bibitem{huber2015glottal}
S.~Huber and A.~Roebel, ``On glottal source shape parameter transformation
  using a novel deterministic and stochastic speech analysis and synthesis
  system,'' in \emph{16th Annual Conference of the International Speech
  Communication Association (INTERSPEECH)}.\hskip 1em plus 0.5em minus
  0.4em\relax ISCA, 2015.

\bibitem{molina2014parametric}
E.~Molina, I.~Barbancho, A.~M. Barbancho, and L.~J. Tard{\'o}n, ``Parametric
  model of spectral envelope to synthesize realistic intensity variations in
  singing voice,'' in \emph{2014 IEEE International Conference on Acoustics,
  Speech and Signal Processing (ICASSP)}.\hskip 1em plus 0.5em minus
  0.4em\relax IEEE, 2014, pp. 634--638.

\bibitem{gentilucci2018vocal}
M.~Gentilucci, L.~Ardaillon, and M.~Liuni, ``Vocal distortion and real-time
  processing of roughness,'' in \emph{International Computer Music Conference
  (ICMC)}, 2018.

\bibitem{gentilucci2019composing}
------, ``Composing vocal distortion: A tool for real-time generation of
  roughness,'' \emph{Computer Music Journal}, vol.~42, no.~4, pp. 26--40, 2019.

\bibitem{qian2019autovc}
K.~Qian, Y.~Zhang, S.~Chang, X.~Yang, and M.~Hasegawa-Johnson, ``Autovc:
  Zero-shot voice style transfer with only autoencoder loss,'' in
  \emph{International Conference on Machine Learning (ICML)}.\hskip 1em plus
  0.5em minus 0.4em\relax PMLR, 2019.

\bibitem{qian2020f0}
K.~Qian, Z.~Jin, M.~Hasegawa-Johnson, and G.~J. Mysore, ``F0-consistent
  many-to-many non-parallel voice conversion via conditional autoencoder,'' in
  \emph{International Conference on Acoustics, Speech and Signal Processing
  (ICASSP)}.\hskip 1em plus 0.5em minus 0.4em\relax IEEE, 2020.

\bibitem{bous2022bottleneck}
F.~Bous and A.~Roebel, ``A bottleneck auto-encoder for f0 transformations on
  speech and singing voice,'' \emph{Information}, vol.~13, no.~3, p. 102, 2022.

\bibitem{tsirulnik2019singing}
L.~Tsirulnik and S.~Dubnov, ``Singing voice database,'' in \emph{International
  Conference on Speech and Computer (ICSC)}.\hskip 1em plus 0.5em minus
  0.4em\relax Springer, 2019.

\bibitem{duan2013nus}
Z.~Duan, H.~Fang, B.~Li, K.~C. Sim, and Y.~Wang, ``The nus sung and spoken
  lyrics corpus: A quantitative comparison of singing and speech,'' in
  \emph{6th Asia-Pacific Signal and Information Processing Association Annual
  Summit and Conference (APSIPA ASC)}.\hskip 1em plus 0.5em minus 0.4em\relax
  IEEE, 2013.

\bibitem{grammalidis2016treasures}
N.~Grammalidis, K.~Dimitropoulos, F.~Tsalakanidou, A.~Kitsikidis, P.~Roussel,
  B.~Denby, P.~Chawah, L.~Buchman, S.~Dupont, S.~Laraba \emph{et~al.}, ``The
  i-treasures intangible cultural heritage dataset,'' in \emph{3rd
  International Symposium on Movement and Computing (MOCO)}, 2016.

\bibitem{koguchi2020pjs}
J.~Koguchi, S.~Takamichi, and M.~Morise, ``Pjs: phoneme-balanced japanese
  singing-voice corpus,'' in \emph{Asia-Pacific Signal and Information
  Processing Association Annual Summit and Conference (APSIPA ASC)}.\hskip 1em
  plus 0.5em minus 0.4em\relax IEEE, 2020.

\bibitem{tamaru2020jvs}
H.~Tamaru, S.~Takamichi, N.~Tanji, and H.~Saruwatari, ``Jvs-music: Japanese
  multispeaker singing-voice corpus,'' \emph{arXiv preprint arXiv:2001.07044},
  2020.

\bibitem{ogawa2021tohoku}
I.~Ogawa and M.~Morise, ``Tohoku kiritan singing database: A singing database
  for statistical parametric singing synthesis using japanese pop songs,''
  \emph{Acoustical Science and Technology}, vol.~42, no.~3, pp. 140--145, 2021.

\bibitem{wilkins2018vocalset}
J.~Wilkins, P.~Seetharaman, A.~Wahl, and B.~Pardo, ``Vocalset: A singing voice
  dataset,'' in \emph{International Society for Music Information Retrieval
  Conference (ISMIR)}.\hskip 1em plus 0.5em minus 0.4em\relax ISMIR, 2018.

\bibitem{ardaillon2017synthesis}
L.~Ardaillon, ``Synthesis and expressive transformation of singing voice,''
  Ph.D. dissertation, Université Pierre et Marie Curie, 2017,
  \url{https://hal.archives-ouvertes.fr/tel-01710926/document}.

\bibitem{glasberg2002model}
B.~R. Glasberg and B.~C. Moore, ``A model of loudness applicable to
  time-varying sounds,'' \emph{Journal of the Audio Engineering Society},
  vol.~50, no.~5, pp. 331--342, 2002.

\bibitem{kingma2014adam}
D.~P. Kingma and J.~Ba, ``Adam: A method for stochastic optimization,'' in
  \emph{3rd International Conference on Learning Representations (ICLR)}, 2014.

\bibitem{ardaillon2019fully}
L.~Ardaillon and A.~Roebel, ``Fully-convolutional network for pitch estimation
  of speech signals,'' in \emph{20th Annual Conference of the International
  Speech Communication Association (INTERSPEECH)}.\hskip 1em plus 0.5em minus
  0.4em\relax ISCA, 2019.

\bibitem{roebel2022neural}
A.~Roebel and F.~Bous, ``Neural vocoding for singing and speaking voices with
  the multi-band excited wavenet,'' \emph{Information}, vol.~13, no.~3, p. 103,
  2022.

\end{thebibliography}

\end{document}